\documentclass[aps,twocolumn,a4paper,showkeys,showpacs,superscriptaddress]{revtex4-1}
\usepackage[dvipdfmx]{graphicx}
\usepackage{amsmath}
\usepackage{bm}
\usepackage{wrapfig}
\usepackage[dvipdfmx]{hyperref}
\hypersetup{
setpagesize=false,
 bookmarksnumbered=true,%
 bookmarksopen=true,%
 colorlinks=true,%
 linkcolor=blue,
 citecolor=blue,
 urlcolor=black,
}

\begin{document}
\title{Search for a stochastic gravitational wave background at 1-5 Hz \\ with Torsion-bar Antenna}


\author{Yuya Kuwahara}
\email{kuwahara@granite.phys.s.u-tokyo.ac.jp}
\affiliation{Department of Physics, University of Tokyo, 7-3-1 Hongo, Bunkyo-ku, Tokyo 113-0033, Japan}

\author{Ayaka Shoda}
\affiliation{Gravitational Wave Project Office, Optical and Infrared Astronomy Division, National Astronomical Observatory, Osawa 2-21-1, Mitaka, Tokyo 181-8588, Japan}

\author{Kazunari Eda}
\affiliation{Department of Physics, University of Tokyo, 7-3-1 Hongo, Bunkyo-ku, Tokyo 113-0033, Japan}
\affiliation{Research center for the early universe (RESCEU), Graduate School of Science, University of Tokyo, Tokyo 113-0033, Japan}

\author{Masaki Ando}
\affiliation{Department of Physics, University of Tokyo, 7-3-1 Hongo, Bunkyo-ku, Tokyo 113-0033, Japan}
\affiliation{Gravitational Wave Project Office, Optical and Infrared Astronomy Division, National Astronomical Observatory, Osawa 2-21-1, Mitaka, Tokyo 181-8588, Japan}
\affiliation{Research center for the early universe (RESCEU), Graduate School of Science, University of Tokyo, Tokyo 113-0033, Japan}


\begin{abstract}
We set the first upper limit on the stochastic gravitational wave (GW) background in the frequency range of 
$1-5\,\mathrm{Hz}$ using a Torsion-bar Antenna (TOBA).
A TOBA is a GW detector designed for the detection of low frequency GWs on the ground, 
with two orthogonal test masses rotated by the incident GWs.
We performed a 24-hour observation run using the TOBA and set upper limits, 
based on frequentist statistics and Bayesian statistics.
The most stringent values are $\Omega_\mathrm{gw}h_0^2 \leq 6.0 \times10^{18}$ (frequentist) and
$\Omega_\mathrm{gw}h_0^2 \leq 1.2 \times 10^{20}$ (Bayesian) both at $2.58\,\mathrm{Hz}$,
where $h_0$ is the Hubble constant in units of $100\,\mathrm{km/s/Mpc}$ 
and $\Omega_\mathrm{gw}$ is the GW energy density per logarithmic frequency interval in units of the closure density.
\end{abstract}

\pacs{04.80.Nn, 95.55.Ym}

\maketitle


\section{Introduction}
Recently, gravitational waves (GWs) were directly detected by LIGO
as the first event GW150914 from a binary black hole merger \cite{abbott2016a}
and subsequently as the second event GW151226 \cite{abbott2016b}.
These discoveries open GW astronomy and thus attract more attention 
not only to binary black holes but also to other GW sources.

A stochastic gravitational wave background (SGWB) is one of the most interesting targets of GWs.
Its origin can be divided into a cosmological one or an astrophysical one.
The former is the isotropic primordial GW produced in the very early universe and
carries to us information that is unavailable by light.
The latter is the superposition of a large number of unresolved sources such as binary black holes
\cite{abbott2016c, Inayoshi:2016hco}
and contains valuable information for astrophysics.

To date, a number of observations have been performed to set the upper limits on the SGWB.
Big-bang nucleosynthesis (BBN) \cite{maggiore2000} 
and the cosmic microwave background (CMB) and matter power spectra \cite{smith2006} 
constrained the cosmological SGWB integrated over all frequencies.
Since these results have no information about the frequency dependence and do not contain the astrophysical SGWB,
it is also necessary to search the SGWB at each frequency band.
In the low frequency ranges (below $1\,\mathrm{mHz}$), the upper limit was set by COBE \cite{allen1996},
pulsar timing \cite{maggiore2000} and Doppler tracking of the Cassini spacecraft \cite{armstrong2003}.
In the middle frequency ranges ($1\,\mathrm{mHz}-1\,\mathrm{Hz}$),
 the upper limit was set by Earth's normal mode oscillation \cite{coughlin2014a},
seismic measurements of the Earth and the Moon  \cite{coughlin2014b, coughlin2014c},
GPS \cite{aoyama2014}, and  Torsion-bar Antennas (TOBAs) \cite{ishidoshiro2011, shoda2014}.
In the high frequency ranges (above $41.5\,\mathrm{Hz}$), 
experiments were performed by LIGO and Virgo \cite{aasi2014},
two LIGO Hanford detectors (H1 and H2) \cite{aasi2015},
cryogenic resonant bars \cite{astone1999},
and a pair of synchronous interferometers \cite{akutsu2008}.
In the frequency range of $1-41.5\,\mathrm{Hz}$, however, the SGWB has yet to be searched 
mainly because of the difficulty of seismic vibration isolation for ground-based detectors.  

In this paper, we report on the first search for the SGWB at $1-5\,\mathrm{Hz}$ using the observation data of 
our upgraded TOBA \cite{shodaDthesis, eda2014}, which, compared with our previous TOBA \cite{ishidoshiro2011},
has improved the seismic vibration isolation at around $1\,\mathrm{Hz}$ by active and passive isolation systems.

 \section{TOBA}
 
 \begin{figure}
\centering
  \includegraphics[width=0.8\hsize]{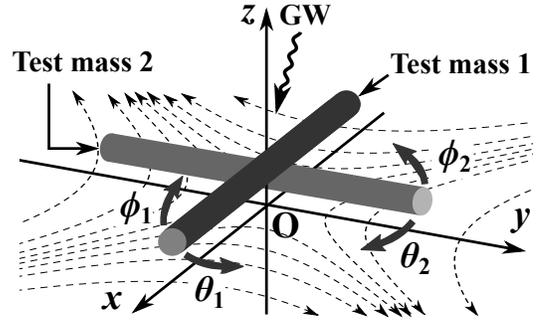}
  \caption{Principle of a TOBA. 
  Two orthogonal test masses are rotated differentially by the tidal force (dotted lines) of the incident GW (wavy line).
  In the case of the GW along the z-axis, 
  the horizontal rotation angle $\theta_1(=-\theta_2)$ is proportional to the GW amplitude 
  while the vertical rotation angles $\phi_1=\phi_2=0$.}
\label{toba}
 \end{figure}

A TOBA \cite{ando2010} is a ground-based GW detector designed for the detection of low-frequency GWs
with two orthogonal test masses, which are rotated by the tidal force of the incident GWs (see Fig. \ref{toba}).
The angular fluctuation $\theta$ of the test masses obeys the equation of motion \cite{ando2010} :
\begin{equation}
I\ddot{\theta}+\gamma \dot{\theta}+\kappa \theta=\frac{1}{4}\ddot{h}_{ij}q^{ij},
\end{equation}
where $I$ is the moment of inertia, $\gamma$ is the damping constant, $\kappa$ is the spring constant,
$h_{ij}$ is the amplitude of the GW, and $q^{ij}$ is the quadrupole moment of the test mass.
In Fourier space, this equation can be reduced to 
\begin{equation}
\tilde{\theta}(f)=\tilde{h}_{ij}(f)q^{ij}/2I,
\end{equation}
above the rotational resonant frequency $f_0=\sqrt{\kappa/I}/2\pi$, 
where a tilde denotes the Fourier amplitude.
Because the resonant frequency of torsion pendulum can be on the order of $1\,\mathrm{mHz}$, 
a TOBA fundamentally has sensitivity in the low frequency ranges (above $1\,\mathrm{mHz}$).
Bisides, the low resonant frequency allows us to easily suppress the effect of the rotational seismic vibration,
which is considered to be originally small.

We have reported experimental results on TOBA in previous papers.
The first prototype composed of a single $20\,\mathrm{cm}$ test mass was constructed 
for its principle verification \cite{ishidoshiro2011}.
The sensitivity was at the level of $10^{-9}\,\mathrm{Hz}^{-1/2}$ in the frequency range of $0.1-1\,\mathrm{Hz}$,
which was limited by seismic noise coupling (above $0.1\,\mathrm{Hz}$) and magnetic noise (below $0.1\,\mathrm{Hz}$).
In order to reduce the seismic coupling, active and passive vibration isolation systems were introduced into
the upgraded TOBA (Phase-II TOBA) \cite{shodaDthesis, shoda2016}, 
leading to the improvement of the sensitivity mainly at around $1\,\mathrm{Hz}$.
In addition, the Phase-II TOBA has three independent outputs of 
one horizontal rotation $\theta \equiv (\theta_1-\theta_2)/2$ and two vertical rotations $\phi_1$ and $\phi_2$,
where the indices 1 and 2 stand for the test mass $1$ and test mass $2$, respectively.
This multi-output system improves the angular resolution 
for short-duration GW signals in the case of a single detector \cite{eda2014}.
The Phase-II TOBA is composed of two $24\,\mathrm{cm}$ test masses, 
with the rotational resonant frequencies of $0.1\,\mathrm{Hz}$.
Displacements at the edges of the test masses are monitored by interferometeric sensors.
In order to keep the sensors within their linear ranges, 
the test masses are feedback controlled by using coil-magnet actuators.

\begin{figure}
\centering
  \includegraphics[width=\hsize]{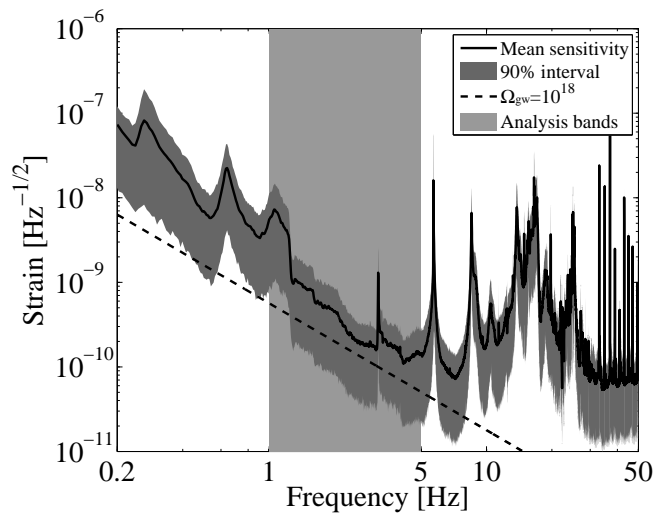}
  \caption{Observed spectral density of GW equivalent strain amplitude by the upgraded TOBA.
  The solid line is the mean sensitivity and the gray region includes $90\%$ data.
  The dashed line is the strain level corresponding to $\Omega_{\mathrm{gw}}h_0^2=10^{18}$.
  The light gray region is the analysis frequency band.}
\label{strain}
 \end{figure}

Using the Phase-II TOBA, we performed a 24-hour observation run from 8:50 UTC, 
December 10, 2014 to 8:50 UTC, December 11, 2014, in Tokyo (35$^\circ$42'49.0"N, 139$^\circ$45'47.0"E).
The spectral density of GW equivalent strain amplitude that is derived from the horizontal rotation is shown in Fig. \ref{strain}.
The sensitivity was at the level of $10^{-10}\,\mathrm{Hz}^{-1/2}$ in most of the range between $1-10\,\mathrm{Hz}$.
The spectral densities of the vertical rotations are not shown, since their sensitivities are two orders of magnitude larger 
than that of the horizontal \cite{shodaDthesis}.
Thus, only horizontal data are used for our analysis below.

\section{analysis}
A target of this work is to set an upper limit on the SGWB.
Searches for the SGWB have often been performed by using a cross-correlation analysis 
with several detectors \cite{allen1999}, which allows the extraction of the signal of the SGWB 
out of the much larger noise background of the detectors.
In our case, however, the cross-correlation analysis is unable to be adopted since we have only one detector.
Still, setting the upper limit is possible without distinguishing the signal and the noise,
which is valid even if all of the data were derived from the SGWB.

The energy density spectrum of the SGWB $\Omega_{\mathrm{gw}}(f)$ is defined as \cite{carr1980}
\begin{equation}
\Omega_{\mathrm{gw}}(f) \equiv \frac{1}{\rho_{\mathrm{c}}}\frac{d\rho_{\mathrm{gw}}}{d \ln f},
\end{equation}
where $d\rho_{\mathrm{gw}}$ is the energy density contained in the frequency interval $df$
and $\rho_{\mathrm{c}} \equiv 3 c^2 H_0^2/8\pi G$ is the critical energy density required to close the universe.
In the definition of $\rho_\mathrm{c}$, $c$ is the speed of light, $H_0$ is the Hubble constant, 
and $G$ is the gravitational constant.
Assuming that the SGWB is isotropic, unpolarized, stationary and Gaussian, 
$\Omega_{\mathrm{gw}}$ is related to the observed GW strain amplitude $\tilde{h}(f)$ 
\cite{ishidoshiro2011, allen1999}:
\begin{equation}
\Omega_{\mathrm{gw}}(f) = \frac{10\pi^2}{3H_0^2}f^3|\tilde{h}(f)|^2,
\label{SGWB}
\end{equation}
where the effect of antenna pattern function of the TOBA is taken into account.
In the following, $\Omega_{\mathrm{gw}}h_0^2 $ is used instead of $ \Omega_{\mathrm{gw}}$ for convenience
because the former is independent of the actual Hubble constant, 
where $h_0 \equiv H_0/(100\,\mathrm{km/s/Mpc})$ is the normalized Hubble constant.

The analysis frequency range was chosen as $1-5\,\mathrm{Hz}$ 
since this was a part of the most sensitive frequency ranges to $\Omega_{\mathrm{gw}}(f)$ (see Fig. \ref{strain}).
The smallest value of mean GW energy density was $\Omega_{\mathrm{gw}}{h_0}^2=2.23\times 10^{18}$ 
at $2.58\,\mathrm{Hz}$.

The calibration to $\Omega_{\mathrm{gw}}$ was done as follows.
The recorded 24-hour raw data $s(t)$, which were error signals in the control systems, 
were divided into $1349$ segments $s_i(t)$ 
of $128\,\mathrm{s}$ with $50\%$ overlap, where $i$  denotes the $i^\mathrm{th}$ segment.
This segment length was chosen so that higher frequency resolution ($8\,\mathrm{mHz}$)  
and sufficient statistics could be obtained in the analysis frequency band, following Ref \cite{ishidoshiro2011}. 
Each segment $s_i(t)$ was independently Fourier transformed into $\tilde{s}_i(f)$ by the use of the fast-Fourier-transform.
Then it was converted into the GW equivalent strain amplitude $\tilde{h}_i(f)=\tilde{s}_i(f) \times (1+G)/MI$.
Here $G$ is the open loop transfer function of the control loop, 
$M$ is the transfer function estimated from mass shape
from the GW amplitude to the displacement at the sensing point, 
and $I$ is the transfer function of the sensor from the displacement to the voltage.
The time series of the energy density of SGWB $\Omega_{\mathrm{gw},\,i}(f)$ was obtained according to Eq.(\ref{SGWB}).

\begin{figure}
\centering
  \includegraphics[width=\hsize]{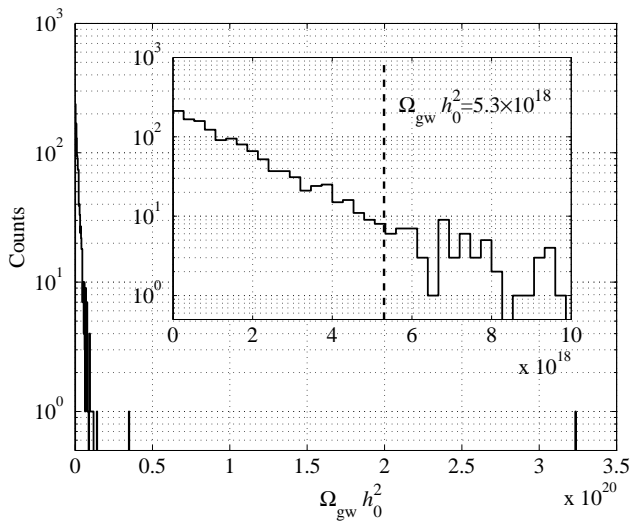}
  \caption{Histogram of $\Omega_{\mathrm{gw}}h_0^2$ at $2.58\,\mathrm{Hz}$. 
  The inset shows an expanded region.
  The dashed line is $\Omega_{\mathrm{gw}}h_0^2=5.3\times10^{18}$, below which $95\%$ of the data are contained.}
\label{frequentist}
 \end{figure}

\begin{figure}
\centering
  \includegraphics[width=\hsize]{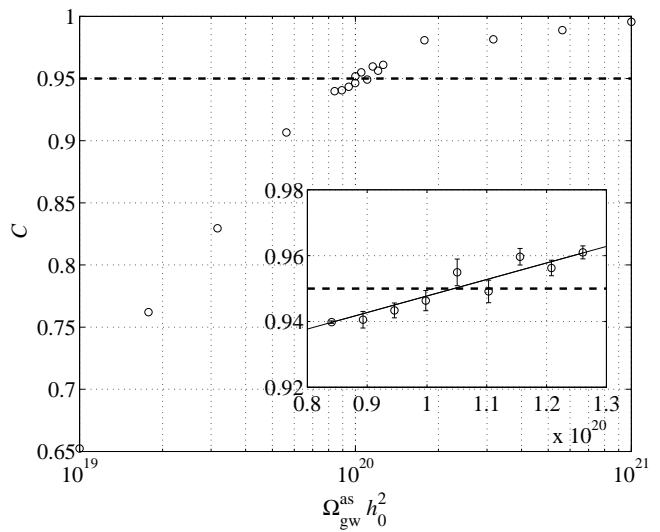}
  \caption{Rate $C$ at $2.58\,\mathrm{Hz}$.
 The inset shows an expanded region near $C=0.95$.
 The solid line is the result of a least-squares fit to the data that are included only within the range of the inset.
 As a result, we obtain the Bayesian upper limit of 
 $\Omega_\mathrm{gw}^\mathrm{B}h_0^2=1.04_{-0.01}^{+0.02}\times10^{20}$.}
\label{baysian}
 \end{figure}

We set two types of upper limit on $\Omega_{\mathrm{gw}}$. 
One is the upper limit based on the frequentist probability.
At each frequency, this  upper limit $\Omega_\mathrm{gw}^\mathrm{F}$ at the $95\%$ confidence level is determined as the value below which $95\%$ of the data are contained:
\begin{equation}
\int_0^{\Omega_\mathrm{gw}^\mathrm{F}}P(\Omega_{\mathrm{gw}})\,d\Omega_{\mathrm{gw}}=0.95,
\end{equation}
where $P(\Omega_{\mathrm{gw}})$ is the probability distribution of $\Omega_{\mathrm{gw},\,i}$.
At $2.58\,\mathrm{Hz}$ where the most stringent value is obtained, for example, the distribution is shown in Fig. \ref{frequentist}.
Since the number of segment is $1349$, the $95\%$ point corresponds to the $(1282\pm8)$-th value of 
$\Omega_\mathrm{gw}^\mathrm{F}h_0^2 =5.3_{-0.3}^{+0.2}\times 10^{18}$.
The error is the $1\,\mathrm{\sigma}$ standard deviation.

The other is the upper limit based on the Bayesian statistics ($\Omega_\mathrm{gw}^\mathrm{B}$).
In this analysis, we derive to what extent the observation data could explain the distributions 
in case a certain value $\Omega^{\mathrm{as}}_{\mathrm{gw}}$ is assumed.
The rate $C$ at which the data cannot explain the assumed distributions is defined as
 \begin{equation}
C(\Omega^{\mathrm{as}}_{\mathrm{gw}}) \equiv \int_{\Omega_{\mathrm{gw}}^{\mathrm{th}}}^{\infty}
Q(\Omega_{\mathrm{gw}},{\mathrm{\Omega_{\mathrm{gw}}^{\mathrm{as}}}})\,d\Omega_{\mathrm{gw}},
\end{equation}
where $\Omega_{\mathrm{gw}}^{\mathrm{th}}$ is the threshold determined by the observation data distribution and 
$Q(\Omega_{\mathrm{gw}},{\mathrm{\Omega_{\mathrm{gw}}^{\mathrm{as}}}})$ is the probability distribution of 
$\Omega_{\mathrm{gw}}$ 
if the SGWB with the mean value of $\Omega_{\mathrm{gw}}=\Omega_{\mathrm{gw}}^{\mathrm{as}}$ exists.
Then the Bayesian upper limit at $95\%$ confidence level is obtained as the assumed value of 
$\Omega_\mathrm{gw}^\mathrm{B}=\Omega_\mathrm{gw}^\mathrm{as}(C=0.95)$.
Although there is arbitrariness in the choice of $\Omega_{\mathrm{gw}}^{\mathrm{th}}$, we adopted 
the $95\%$ point of the observation data distribution, which is the same as the frequentist upper limit.

$Q(\Omega_{\mathrm{gw}},{\mathrm{\Omega_{\mathrm{gw}}^{\mathrm{as}}}})$ is obtained by the signal injection 
into the observation data. 
First, Gaussian noises that have the same length as the observation data are made, divided into segments and 
Fourier transformed in the same manner as the analysis of the observation data. 
Each segment is normalized so that the mean power of the noise is $\Omega_{\mathrm{gw}}^{\mathrm{as}}$.
Then it is converted into strain, adding to each segment of the observation strain data.
The distribution of the resulting data is $Q(\Omega_{\mathrm{gw}},{\mathrm{\Omega_{\mathrm{gw}}^{\mathrm{as}}}})$.

At each frequency, we changed the injection value of $\Omega_{\mathrm{gw}}^{\mathrm{as}}$ 
 and calculated $C$ for five times.
At $2.58\,\mathrm{Hz}$, for example, the result is shown in Fig. \ref{baysian}.
The data points and the error bars show the averages and the standard deviations, respectively.
In this case, the Bayesian upper limit is $\Omega_\mathrm{gw}^\mathrm{B}h_0^2=1.04_{-0.01}^{+0.02}\times10^{20}$.
This error comes from the statistical (fitting) error.

The systematic errors arise from the calibration.
The main errors are the uncertainties of the efficiencies of the sensors, which is $7.3\%$.
Including other small fractions and assuming that they are independent, 
the total systematic errors are estimated to be $8.9\%$.

\begin{figure}
\centering
  \includegraphics[width=0.99\hsize]{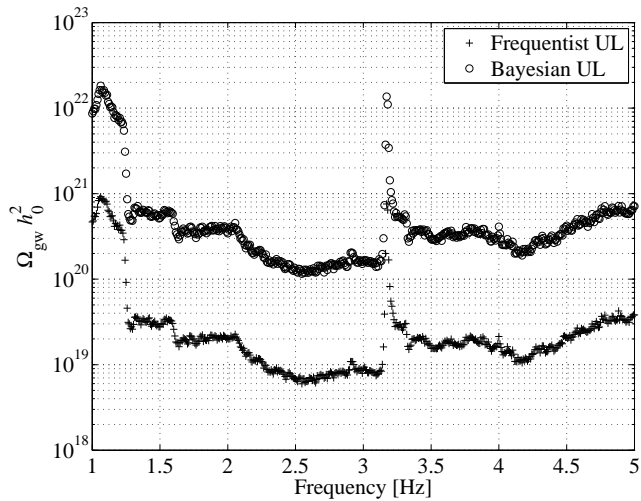}
  \caption{Upper limits of $\Omega_\mathrm{gw}h_0^2$ at the $95\%$ confidence level.
  The most stringent constraints are $\Omega_\mathrm{gw}^\mathrm{F}h_0^2=6.0\times10^{18}$ and 
  $\Omega_\mathrm{gw}^\mathrm{B}h_0^2=1.2\times10^{20}$ both at $2.58\,\mathrm{Hz}$. }
\label{UL_list_table}
 \end{figure}

Taking the statistical errors and the systematic errors into account, 
we finally obtained conservative upper limits, 
which means that positive signs are adopted for both errors.
The results are shown in Fig. \ref{UL_list_table}.
At $2.58\,\mathrm{Hz}$, for example, the final upper limits are 
$\Omega_\mathrm{gw}^\mathrm{F}h_0^2=6.0\times10^{18}$ and 
$\Omega_\mathrm{gw}^\mathrm{B}h_0^2=1.2\times10^{20}$.

\begin{figure}
\centering
  \includegraphics[width=0.99\hsize]{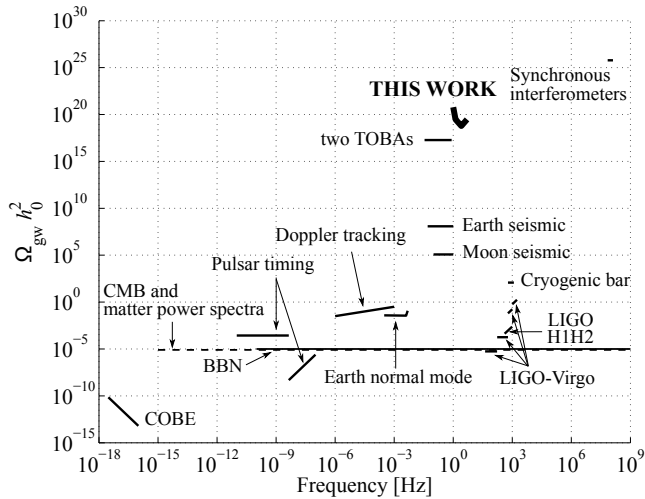}
  \caption{Current upper limits on the energy density of the SGWB.
The bold line is our new upper limit (frequentist).}
\label{UL_list}
 \end{figure}

\section{discussion}
 
Our new results and the current upper limits in other frequency bands are shown in Fig. \ref{UL_list}.
Compared with the upper limits below $1\,\mathrm{Hz}$ next to our analysis bands, 
the results of the Phase-II TOBA are much greater than that of the seismic measurement of the moon, 
which is the most stringent upper limit at $0.1-1\,\mathrm{Hz}$.
However, as described in Ref. \cite{coughlin2014c}, 
further search for the SGWB by using such seismic mesurements is difficult
unless seismometers are set on other quieter planets than the moon.
On the other hand, a TOBA has a potential to be further upgraded.
The sensitivity of $10^{-19}\,\mathrm{/\sqrt{Hz}}$ at $0.1-1\,\mathrm{Hz}$ will be realized 
by the final configuration of a TOBA with $10\,\mathrm{m}$-scale bars \cite{ando2010}.
It is expected to be able to search the SGWB beyond the BBN limit $\Omega_{\mathrm{gw}} \sim 10^{-5}$,
with a one-year cross-correlation analysis by a pair of two final TOBAs.

In our results, $\Omega_\mathrm{gw}^\mathrm{B}h_0^2$ are roughly $20$ times greater than 
$\Omega_\mathrm{gw}^\mathrm{F}h_0^2$ although both analyses should lead to essentially similar results.
This is because we have a single detector and a single data set.
In this case, the output of the detector and the injection signals are indistinguishable.
Therefore to extract the injection signals from the summed data, the injection signals are inevitably greater than 
the original detector's output.

\section{conclusion}
We performed the search for the SGWB using the observation data of the Phase-II TOBA. 
As a result, we obtained the first upper limits between $1-5\,\mathrm{Hz}$ on the SGWB at the $95\%$ confidence level.
The most stringent values are $\Omega_\mathrm{gw}^\mathrm{F}h_0^2=6.0\times10^{18}$ (frequentist)
and $\Omega_\mathrm{gw}^\mathrm{B}h_0^2 = 1.2 \times 10^{20}$ (Bayesian) both at $2.58\,\mathrm{Hz}$.\\*

This work was supported by JSPS KAKENHI Grants No. 24244031 (M. A.),
JSPS Fellows Grant No. 24.7531 (A. S.),
and JSPS Fellows Grant No. 26.8636 (K. E.).

\end{document}